\documentclass[iop]{emulateapj}  

\usepackage{ulem}

\newcommand\aov{\ifmmode{\alpha_{\rm ov}}\else $\alpha_{\rm ov}$\fi}
\newcommand\fov{\ifmmode{f_{\rm ov}}\else $f_{\rm ov}$\fi}
\newcommand\amix{\ifmmode{\alpha_{\rm MLT}}\else $\alpha_{\rm MLT}$\fi}
\newcommand\gs{GS98}
\newcommand\as{A09}

\shortauthors{Claret \& Torres}
\shorttitle{Core overshooting}

\begin{document} 

\submitted{Accepted for publication in The Astrophysical Journal}

\title{The dependence of convective core overshooting on stellar mass:
  semi-empirical determination using the diffusive approach with two
  different element mixtures}

\author{
Antonio Claret\altaffilmark{1} and
Guillermo Torres\altaffilmark{2}
}

\altaffiltext{1}{Instituto de Astrof\'{\i}sica de Andaluc\'{\i}a,
  CSIC, Apartado 3004, 18080 Granada, Spain}

\altaffiltext{2}{Harvard-Smithsonian Center for Astrophysics, 60
  Garden St., Cambridge, MA 02138, USA}

\begin{abstract}
Convective core overshooting has a strong influence on the evolution
of stars of moderate and high mass. Studies of double-lined eclipsing
binaries and stellar oscillations have renewed interest in the
possible dependence of overshooting on stellar mass, which has been
poorly constrained by observations so far.  Here we have used a sample
of 29 well-studied double-lined eclipsing binaries in key locations of
the H-R diagram to establish the explicit dependence of \fov\ on mass,
where \fov\ is the free parameter in the diffusive approximation to
overshooting. Measurements of the masses, radii, and temperatures of
the binary components were compared against stellar evolution
calculations based on the MESA code to infer semi-empirical values of
\fov\ for each component.  We find a clear mass dependence such that
\fov\ rises sharply from zero in the range 1.2--$2.0~M_{\sun}$, and
levels off thereafter up to the 4.4~$M_{\sun }$ limit of our sample.
Tests with two different element mixtures indicate the trend is the
same, and we find it is also qualitatively similar to the one
established in our previous study with the classical step-function
implementation of overshooting characterized by the free parameter
\aov. Based on these measurements we infer an approximate relationship
between the two overshooting parameters of $\aov/\fov = 11.36 \pm
0.22$, with a possible dependence on stellar properties.
\end{abstract}

\keywords{stars: evolution; 
stars: interiors;  
stars: core overshooting;  
stars: eclipsing binaries}

\section{Introduction}
\label{sec:introduction}

Theoretical research on the extent of convective cores in stars dates
back to the beginning of the 1960's. Convective elements that travel
past the boundary of the core as defined by the classical
Schwarzschild criterion, and extend its size, give rise to the
phenomenon typically referred to as convective core overshooting.
The extra mixing has a strong influence on the evolution of stars. It
extends their main-sequence lifetimes, and increases the degree of
mass concentration towards the center. The latter property can be
measured directly by determining the rate of apsidal motion in close,
eccentric binary systems \citep[see, e.g.,][]{Claret:10}. The
bibliography on the subject of convective core overshooting is quite
extensive, and for the most relevant sources of interest for the
present work we refer the reader to our earlier paper on the topic
\citep[][hereafter Paper~I]{CT:16}.

While it is now widely accepted that core overshooting is needed in
order to explain the observations of stars in a variety of stellar
populations (see the discussion in the aforementioned paper), the
degree to which overshooting might depend on stellar mass has been
much less clear. Many (though not all) current grids of stellar
evolution models assume some form of mass dependence for the extra
mixing, but those prescriptions are largely ad hoc. Empirical support
has been scant, and sometimes contradictory.  The issue has commonly
been addressed most directly using measurements of double-lined
eclipsing binaries (DLEBs), and in recent years also using diagnostics
from asteroseismology. In the framework of the classical prescription
for overshooting the extra distance traveled by the convective
elements beyond the limits of the core is represented by $d_{\rm ov} =
\aov\ H_p$, where $H_p$ is the pressure scale height and \aov\ is the
classical overshooting parameter.  Studies using DLEBs have found
results ranging from a strong dependence of the \aov\ on mass
\citep{Ribas:00} to a more uncertain and much less pronounced
dependence \citep{Claret:07}, and even no dependence at all
\citep{Meng:14, Stancliffe:15}. The DLEB samples in these
investigations have typically been small, not exceeding a dozen or so
systems (often far fewer), and have generally been restricted to the
main sequence. Asteroseismic studies of single stars have so far been
performed over limited ranges in mass and evolutionary state, but do
suggest changes in overshooting that are not inconsistent with results
described below \citep[see, e.g.,][]{Aerts:13, Deheuvels:16}.

The most comprehensive observational study of the dependence of core
overshooting on stellar mass was presented in Paper~I, where we made
use of a much larger set of 33 DLEBs in the range of
1.2--$4.4~M_{\odot}$ with accurate determinations of their physical
properties (masses, radii, effective temperatures, and in many cases
metallicities). These systems were carefully selected to be evolved
enough so that we could discern the effects of core overshooting and
infer a possible dependence on the mass of the star. A clear relation
emerged between \aov\ and stellar mass indicating a steep rise to
about $2~M_{\odot}$ with little or no change beyond this mass. No
measurable dependence on metallicity was seen. This semi-empirically
determined pattern of overshooting for stars of different mass has
found an important practical application for stellar evolution models
and has now been incorporated into the new grid of Yale-Potsdam
Stellar Isochrones \citep[YaPSI;][]{Spada:17}.

In recent years an alternate prescription for core overshooting has
become increasingly popular, in which the region of convective
overshooting is not fully mixed, as in the classical method. Rather,
as they travel outwards, the convective elements disintegrate
following a diffusive process and the decay of the turbulent velocity
field is exponential.  Following \cite{Freytag:96} and
\cite{Herwig:97}, the diffusion coefficient in the overshooting region
is given by
\begin{eqnarray}
D_{\rm ov} = D_0\,\exp\left({\frac{-2z}{H_{\nu}}}\right),
\end{eqnarray}
where $D_0$ is the diffusion coefficient at the convective boundary,
$z$ is the geometric distance from the edge of the convective zone,
$H_{\nu}$ is the velocity scale height at the convective boundary
expressed as $H_{\nu} = \fov\ H_p$, and the coefficient \fov\ is a
free parameter governing the width of the overshooting layer.

Comparisons between models using both approximations to overshooting
over limited ranges in mass indicate the resulting internal structures
are rather similar \citep[e.g.,][]{Herwig:97, Noels:10, Magic:10},
suggesting perhaps a rough scaling between \aov\ and \fov. However,
some asteroseismic studies claim to be able to distinguish the
internal structures from both types of models, implying they are in
fact not quite the same.  For example, \cite{Mora:15, Mora:16} find
evidence that the diffusive formulation of overshooting seems better
able to explain the observed pattern of period spacings of slowly
pulsating B stars than the classical step-function formulation.

The ``equivalence'' between \aov\ and \fov\ has yet to be explored
more fully over a wide range of masses and evolutionary states. While
a calibration of \aov\ with stellar mass now exists (Paper~I), the
precise quantitative dependence of \fov\ with mass is not known at the
present time, as has been pointed out by \cite{Magic:10}, although one
might expect it to be at least qualitatively similar. One of the
recent series of publicly available evolutionary models that uses the
diffusive approximation for overshooting is the large grid of MESA
Isochrones and Stellar Tracks (MIST) by \cite{Choi:16}, which is based
on the Modules for Experiments in Stellar Astrophysics package
\citep[MESA;][]{Paxton:11, Paxton:13, Paxton:15}. These models assume
\fov\ is independent of stellar mass, and adopt a fixed value of $\fov
= 0.016$ for all stars with convective cores, tuned to reproduce the
shape of the main-sequence turnoff in the color-magnitude diagram
(CMD) of the open cluster M67. If \fov\ actually varies with mass in
some fashion, as we expect, the impact of using a constant
\fov\ rather than a mass-dependent one may not be immediately obvious
in fitting isochrones to CMDs because of the various degeneracies and
uncertainties involved.\footnote{In an extreme example of this
  degeneracy, \cite{Michaud:04} have shown that the CMD of M67 can be
  well fit even {\it without} overshooting, if diffusion is included
  \citep[see also][]{Viani:17}.} On the other hand, DLEBs with well
measured parameters may provide more sensitive tests.

Without access to an empirical calibration of the mass dependence of
\fov, modelers wishing to use the now more common diffusive
approximation are left with little choice but to adopt a fixed value
of this parameter constrained in some way by observations, as in the
case just described, or to fall back on artificial trends informed
perhaps by theoretical expectations \citep[see also][]{Magic:10}. This
unsatisfactory situation provides the main motivation for our work,
which is to explore the dependence of overshooting on stellar mass
using the diffusive formulation, by means of DLEBs, much in the same
way as we did in Paper~I with the simpler step-function prescription.
Our hope is that the results may provide an empirical basis for a more
realistic implementation of diffusive convective core overshooting in
future models of stellar evolution.  An additional goal is to extend
our earlier work and examine the impact on the results of the element
mixture and the primordial helium abundance, which lead to differences
in the opacities.

The paper is structured as follows: In Section~\ref{sec:sample} we
present our observational sample consisting of 29 well-studied DLEBs.
Section~\ref{sec:methods} describes the new prescription for
overshooting we use here, the stellar evolution code, and our
methodology to derive for each star the semi-empirical values of
\fov\ and the mixing length parameters \amix, as well as the system
metallicity. Our results on the mass dependence of \fov\ are reported
in Section~\ref{sec:results} for two different element mixtures, and a
general discussion of the results is given in
Section~\ref{sec:discussion}.  We summarize our findings in
Section~\ref{sec:conclusions}. Finally, in Appendix~A we use a simple
analytical model based on homology transformations and the
differential equations of stellar structure to investigate the role of
opacities, the equation of state, and nuclear energy generation rates
on the enlargement of the convective core due to overshooting. This
provides important insight into the behavior of \fov\ as a function of
mass reported in Section~\ref{sec:results}.

\section{Observational sample}
\label{sec:sample}

In our earlier study of Paper~I \citep{CT:16} we had assembled a set
of 33 detached eclipsing binary systems with well determined masses
and radii (typically good to 3\% or better) selected so that they are
sufficiently evolved for the effects of overshooting to be noticeable.
Many of the more evolved systems belong to the Large or Small
Magellanic Clouds (LMC, SMC), and are preferentially more massive and
more metal-poor than systems from the solar neighborhood, which tend
to have compositions near solar.  Four of those field binaries,
$\chi^2$~Hya, YZ~Cas, V885~Cyg, and VV~Crv, did not quite meet the
requirement adopted there that the individual ages inferred for the
components from a comparison with stellar evolution models be within
5\% of each other.  They were nevertheless retained in that study
because their mass ratios are very different from unity and therefore
provide stronger leverage for the model fits.

The sample selected for the present work draws heavily on the same set
of binaries, though with a few differences. One of the problematic
systems, V885~Cyg, now provides satisfactory fits to the models used
here within the 5\% cap on the age difference, and has been included.
This is likely the result of the use of a different stellar evolution
code than in Paper~I, as described in the next section.  On the other
hand, for HY~Vir (another system with dissimilar masses) we now find
poor fits to the models employed for this work, and we have therefore
preferred to exclude it. This binary and the three others dropped from
the sample in Paper~I are interesting in their own right, and a full
investigation of the reasons for the poor fits, which may possibly
provide further insights into the models, will be the subject of a
future paper.

The final sample for this work contains 29 DLEBs.  Additionally, in a
few cases with primary and secondary masses that are indistinguishable
within the uncertainties, the use of different models here than in
Paper~I has led to a different conclusion regarding which component of
the system is more evolved.  For the benefit of the reader
Table~\ref{tab:sample} lists the properties of the 29 systems we use
for this analysis, with the binaries arranged in order of decreasing
primary mass and the more evolved star given on the first line.

\begin{deluxetable*}{llllcc}
\tabletypesize{\scriptsize}
\tablewidth{0pc}
\tablecaption{Binaries systems in our sample.\label{tab:sample}}
\tablehead{
\colhead{Name} & 
\colhead{Mass ($M_{\sun}$)} & 
\colhead{Radius ($R_{\sun}$)} & 
\colhead{$T_{\rm eff}$ (K)} & 
\colhead{[Fe/H]} &
\colhead{Source}
}
\startdata
SMC-108.1-14904         &   4.429~$\pm$~0.037   &  64.05~$\pm$~0.50   &  4955~$\pm$~90   & $-$0.80~$\pm$~0.15 &  1  \\ 
                        &   4.416~$\pm$~0.041   &  46.95~$\pm$~0.53   &  5675~$\pm$~105  &                    &     \\ [1ex]
OGLE-LMC-ECL-CEP-0227   &   4.165~$\pm$~0.032   &  34.92~$\pm$~0.34   &  6050~$\pm$~160  &                    &  2  \\
                        &  4.134~$\pm$~0.037    &  44.85~$\pm$~0.29   &  5120~$\pm$~130  &                    &     \\ [1ex]
OGLE-LMC-ECL-06575      &   4.152~$\pm$~0.030   &  39.79~$\pm$~1.35   &  4903~$\pm$~72   & $-$0.45~$\pm$~0.10 &  3  \\
                        &  3.966~$\pm$~0.032    &  49.35~$\pm$~1.45   &  4681~$\pm$~77   &                    &     \\ [1ex]
OGLE-LMC-ECL-CEP-2532   &    3.90~$\pm$~0.10    &  28.95~$\pm$~1.4    &  6345~$\pm$~150  &                    &  4  \\
                        &   3.83~$\pm$~0.10     &   37.7~$\pm$~1.7    &  4800~$\pm$~220  &                    &     \\ [1ex]
LMC-562.05-9009         &    3.70~$\pm$~0.03    &   28.6~$\pm$~0.2    &  6030~$\pm$~150: &                    &  5  \\
                        &   3.60~$\pm$~0.03     &   26.6~$\pm$~0.2    &  6030~$\pm$~150: &                    &     \\ [1ex]
OGLE-LMC-ECL-26122      &   3.593~$\pm$~0.055   &  32.71~$\pm$~0.51   &  4989~$\pm$~80   & $-$0.15~$\pm$~0.10 &  3  \\
                        &  3.411~$\pm$~0.047    &  22.99~$\pm$~0.48   &  4995~$\pm$~81   &                    &     \\ [1ex]
OGLE-LMC-ECL-01866      &  3.574~$\pm$~0.038    &  46.96~$\pm$~0.61   &  4541~$\pm$~85   & $-$0.70~$\pm$~0.10 &  3  \\
                        &   3.575~$\pm$~0.028   &  28.20~$\pm$~1.06   &  5327~$\pm$~72   &                    &     \\ [1ex]
OGLE-SMC-113.3-4007     &   3.561~$\pm$~0.025   &   48.4~$\pm$~0.7    &  4813~$\pm$~100  &                    &  6  \\
                        &  3.504~$\pm$~0.028    &   45.8~$\pm$~0.7    &  4800~$\pm$~100  &                    &     \\ [1ex]
OGLE-LMC-ECL-10567      &   3.345~$\pm$~0.040   &   25.6~$\pm$~1.6    &  5067~$\pm$~73   & $-$0.81~$\pm$~0.20 &  3  \\
                        &  3.183~$\pm$~0.038    &   36.0~$\pm$~2.0    &  4704~$\pm$~80   &                    &     \\ [1ex]
OGLE-LMC-ECL-09144      &   3.303~$\pm$~0.028   &  26.18~$\pm$~0.31   &  5288~$\pm$~81   & $-$0.23~$\pm$~0.10 &  3  \\
                        &  3.208~$\pm$~0.026    &  18.64~$\pm$~0.30   &  5470~$\pm$~96   &                    &     \\ [1ex]
OGLE-051019.64-685812.3 &   3.278~$\pm$~0.032   &  26.05~$\pm$~0.29   &  5300~$\pm$~100  &                    &  7  \\
                        &  3.179~$\pm$~0.029    &  19.76~$\pm$~0.34   &  5450~$\pm$~100  &                    &     \\ [1ex]
OGLE-LMC-ECL-09660      &   2.988~$\pm$~0.018   &  43.87~$\pm$~1.14   &  4677~$\pm$~75   & $-$0.44~$\pm$~0.10 &  3  \\
                        &  2.969~$\pm$~0.020    &  23.75~$\pm$~0.66   &  5352~$\pm$~70   &                    &     \\ [1ex]
SMC-101.8-14077         &   2.835~$\pm$~0.055   &  23.86~$\pm$~0.31   &  5170~$\pm$~90   & $-$1.01~$\pm$~0.15 &  1  \\
                        &  2.725~$\pm$~0.034    &  17.90~$\pm$~0.50   &  5580~$\pm$~95   &                    &     \\ [1ex]
$\alpha$ Aur            &  2.5687~$\pm$~0.0074  &  11.98~$\pm$~0.57   &  4970~$\pm$~50   & $-$0.04~$\pm$~0.06 &  8  \\
                        & 2.4828~$\pm$~0.0067   &   8.83~$\pm$~0.33   &  5730~$\pm$~60   &                    &     \\ [1ex]
WX Cep                  &   2.533~$\pm$~0.050   &  3.996~$\pm$~0.030  &  8150~$\pm$~250  &                    &  7  \\
                        &  2.324~$\pm$~0.045    &  2.712~$\pm$~0.023  &  8900~$\pm$~250  &                    &     \\ [1ex]
V1031 Ori               &   2.468~$\pm$~0.018   &  4.323~$\pm$~0.034  &  7850~$\pm$~500  &                    &  7  \\
                        &  2.281~$\pm$~0.016    &  2.978~$\pm$~0.064  &  8400~$\pm$~500  &                    &     \\ [1ex]
V364 Lac                &   2.333~$\pm$~0.014   &  3.309~$\pm$~0.021  &  8250~$\pm$~150  &                    &  7  \\
                        &  2.295~$\pm$~0.024    &  2.986~$\pm$~0.020  &  8500~$\pm$~150  &                    &     \\ [1ex]
SZ Cen                  &   2.311~$\pm$~0.026   &  4.556~$\pm$~0.032  &  8100~$\pm$~300  &                    &  7  \\
                        &  2.272~$\pm$~0.021    &  3.626~$\pm$~0.026  &  8380~$\pm$~300  &                    &     \\ [1ex]
OGLE-LMC-ECL-25658      &  2.230~$\pm$~0.019    &  27.57~$\pm$~0.24   &  4721~$\pm$~75   & $-$0.63~$\pm$~0.10 &  9  \\
                        &  2.229~$\pm$~0.019    &  21.41~$\pm$~0.15   &  4860~$\pm$~70   &                    &     \\ [1ex]
V885 Cyg                &   2.228~$\pm$~0.026   &  3.387~$\pm$~0.026  &  8150~$\pm$~150  &                    &  7  \\
                        &  2.000~$\pm$~0.029    &  2.346~$\pm$~0.017  &  8375~$\pm$~150  &                    &     \\ [1ex]
AI Hya                  &   2.140~$\pm$~0.038   &  3.916~$\pm$~0.031  &  6700~$\pm$~60   &                    &  7  \\
                        &  1.973~$\pm$~0.036    &  2.767~$\pm$~0.019  &  7100~$\pm$~65   &                    &     \\ [1ex]
AY Cam                  &   1.905~$\pm$~0.040   &  2.772~$\pm$~0.020  &  7250~$\pm$~100  &                    &  7  \\
                        &  1.709~$\pm$~0.036    &  2.026~$\pm$~0.017  &  7395~$\pm$~100  &                    &     \\ [1ex]
SMC-130.5-04296         &  1.854~$\pm$~0.025    &  25.44~$\pm$~0.25   &  4912~$\pm$~80   & $-$0.88~$\pm$~0.15 &  1  \\ 
                        &   1.805~$\pm$~0.027   &  46.00~$\pm$~0.35   &  4515~$\pm$~75   &                    &     \\ [1ex]
OGLE-LMC-ECL-03160      &  1.799~$\pm$~0.028    &  37.42~$\pm$~0.52   &  4490~$\pm$~82   & $-$0.48~$\pm$~0.20 &  3  \\
                        &   1.792~$\pm$~0.027   &  16.36~$\pm$~1.06   &  4954~$\pm$~83   &                    &     \\ [1ex]
EI Cep                  &  1.7716~$\pm$~0.0066  &  2.897~$\pm$~0.048  &  6750~$\pm$~100  &                    &  7  \\
                        & 1.6801~$\pm$~0.0062   &  2.330~$\pm$~0.044  &  6950~$\pm$~100  &                    &     \\ [1ex]
SMC-126.1-00210         &   1.674~$\pm$~0.037   &  43.52~$\pm$~1.02   &  4480~$\pm$~70   & $-$0.86~$\pm$~0.15 &  1  \\
                        &  1.669~$\pm$~0.039    &  39.00~$\pm$~0.98   &  4510~$\pm$~70   &                    &     \\ [1ex]
HD 187669               &   1.505~$\pm$~0.004   &  22.62~$\pm$~0.50   &  4330~$\pm$~70   & $-$0.25~$\pm$~0.10 & 10  \\
                        &  1.504~$\pm$~0.004    &  11.33~$\pm$~0.28   &  4650~$\pm$~80   &                    &     \\ [1ex]
OGLE-LMC-ECL-15260      &  1.426~$\pm$~0.022    &  42.17~$\pm$~0.33   &  4320~$\pm$~81   & $-$0.47~$\pm$~0.15 &  3  \\
                        &   1.440~$\pm$~0.024   &  23.51~$\pm$~0.69   &  4706~$\pm$~87   &                    &     \\ [1ex]
AI Phe                  &  1.2336~$\pm$~0.0045  &  2.932~$\pm$~0.048  &  5010~$\pm$~120  & $-$0.14~$\pm$~0.10 &  7  \\
                        & 1.1934~$\pm$~0.0041   &  1.818~$\pm$~0.024  &  6310~$\pm$~150  &                    &     
\enddata
\tablecomments{
The first line for each system corresponds to the more evolved star.
Temperatures for LMC-562.05-9009 are listed as uncertain in the
original source. The [Fe/H] value adopted here for OGLE-LMC-ECL-25658
is the average of the individual estimates reported. Sources are:
(1) \cite{Graczyk:14}; 
(2) \cite{Pilecki:13}; 
(3) \cite{Pietrzynski:13}; 
(4) \cite{Pilecki:15}; 
(5) \cite{Gieren:15}; 
(6) \cite{Graczyk:12}; 
(7) \cite{Torres:10}; 
(8) \cite{Torres:15}; 
(9) \cite{Elgueta:16};
(10) \cite{Helminiak:15}.
}
\end{deluxetable*}

\section{Stellar models and fitting methods}
\label{sec:methods}

In the classical formalism for convective core overshooting, sometimes
referred to as the ``step-function'' approximation, the region beyond
the core boundary as specified by the Schwarzschild criterion is
assumed to be fully mixed.  This simple prescription has found wide
application in many of the current series of stellar evolution models,
and a calibration of \aov\ as a function of stellar mass was presented
in Paper~I.

For the present work we have explored an alternate, exponentially
decaying formulation for overshooting that is becoming more common,
and is characterized by the free parameter \fov\ described earlier
that determines the width of the overshooting layer. The temperature
gradient in this region is assumed to be radiative, and we adopt equal
\fov\ values above the hydrogen and helium burning regions.  As
mentioned before, recent evidence from an asteroseismic study of the
slowly rotating pulsating B stars KIC~10526294 and KIC~7760680 by
\cite{Mora:15, Mora:16} indicates this prescription appears to be
favored over the classical one, at least in these two
cases.\footnote{With the present list of binaries we are not able to
  determine whether one formalism is better than the other. A sample
  with more direct information on the stellar interiors would be
  needed, as could be supplied by accurate measurements of apsidal
  motion in eccentric systems.}

While our previous work in Paper~I made use of the Granada stellar
evolution code \citep{Claret:04}, the diffusive formulation for
overshooting is not yet implemented in that code, so for the present
study evolutionary tracks were computed instead with the MESA code
\citep{Paxton:11, Paxton:13, Paxton:15}, version 7385. A general
description of the input physics of the MESA code is given in the
above sources.  For stars with convective envelopes we employed the
standard mixing-length formalism \citep{Vitense:58}, where \amix\ is a
free parameter.\footnote{The value of the mixing length parameter for
  the Sun in these models, with the same input physics as used here,
  is $\amix = 1.84$ \citep[see][]{Torres:15}.}  The third-degree
equation relating the temperature gradients was solved using the
Henyey option in MESA. Microscopic diffusion was included, and the
condition to determine the edge of the convection zone is the
Schwarzschild criterion. As we did in our earlier work, our
evolutionary calculations begin at the pre-main-sequence phase. We
have not considered stellar rotation in these models, and mass loss
was taken into account with the prescription by \cite{Reimers:77} with
the efficiency coefficient set to $\eta = 0.2$. We note, though, that
the effects of mass loss in our sample are hardly significant,
amounting to no more than 1\% of the initial mass for our most massive
giants (i.e., at the level of the observational errors in the
masses). The high-temperature opacities were taken from the tables by
\cite{Iglesias:96}, and for lower temperatures we used the tables of
\cite{Ferguson:05}.

In our earlier study we had adopted the element mixture of
\cite{Grevesse:98}, which results in a present-day solar metallicity
of $Z_{\sun} = 0.0189$. The primordial helium content was set to $Y_p
= 0.24$, and the adopted slope for the Galactic enrichment law was $\Delta
Y/\Delta Z = 2.0$. For the present work we have chosen to use the same
element mixture initially to facilitate the comparison, but we also
used the more recent mixture of \cite{Asplund:09}, for which $Z_{\sun}
= 0.0134$, to explore the effect of a change in opacities.  This new
mixture was paired with an enrichment law specified by $Y_p = 0.249$
\citep{Planck:16} and a slope $\Delta Y/\Delta Z = 1.67$. Thus, we
performed two sets of independent calculations:
\begin{itemize}
\item \gs\ set: \cite{Grevesse:98} mixture,\\ $Y_p = 0.24$, $\Delta Y/\Delta Z = 2.0$ (same as Paper~I);
\item \as\ set: \cite{Asplund:09} mixture,\\ $Y_p = 0.249$, $\Delta Y/\Delta Z = 1.67$.
\end{itemize}
These choices allow for two comparisons of interest: (1) to study the
results from the two different overshooting formulations by comparing
the \gs\ set with the results of Paper~I, which used the classical
step-function approximation with same mixture and helium enrichment
law; and (2) to explore the influence of the element mixture and the
helium content on overshooting using the same (diffusive)
prescription for the phenomenon.

The fitting procedure applied here to infer the amount of overshooting
for each star and the optimal abundance for each system is similar to
the one used in Paper~I, as described below. The observational
constraints are the measured masses ($M$), radii ($R$), and effective
temperatures ($T_{\rm eff}$). The best match was sought between these
observations and evolutionary tracks calculated for the measured
masses of each component, allowing \fov\ and \amix\ to vary freely for
each star, and letting $Z$ also be variable but assumed to be the same
for the two stars. As only about half of our systems have a measured
chemical composition, we have preferred to use that information as a
consistency check rather than a strict constraint.

For this work we have again opted for the mixed approach applied in
our earlier study that combines a relatively coarse grid search of
parameter space with manual fine tuning to arrive at the final
best-fit values. The manual adjustment has the advantage over a far
more computationally expensive fine grid search that the solutions are
easily inspected at each step of the process to avoid inconsistencies
in the location of the components in the temperature-radius
diagram. This is particularly valuable in rapid phases of evolution
that could lead to unphysical situations (e.g., the more massive star
being less evolved).  For Paper~I we had computed a grid of
evolutionary tracks spanning a range of \aov\ values from 0.00 to 0.40
in steps of 0.05, and \amix\ values between 1.0 and 2.0 (extended in
some cases up to 2.7) with a step of 0.1. We have reused this grid
here, not for actual fits but only in a differential sense, to gauge
the direction in which predicted stellar parameters (temperature,
radius, age) change as the overshooting or mixing length are modified,
and aid our manual search for better fits. As the relevant variable
for this work is \fov\ rather than \aov, we assumed a rough scaling
between the two (see also Section~\ref{sec:comparison}) such that
$\aov/\fov \approx 10$ \citep[e.g.,][]{Herwig:97, Noels:10}, and used
it to re-label our coarse grid in terms of \fov, which we then used as
indicated above to guide our solutions.  Based on this information and
initial estimates of \fov, \amix, and $Z$ we computed finer grids
of MESA models with the diffusive overshooting approximation over
shorter ranges tailored to each system. Small manual adjustments to
the parameters were made in the final approach to the optimal fits
until the predicted effective temperatures and radii agreed with the
measured values approximately within their uncertainties, requiring
also that the binary components have a similar age. To account for
imperfections in the models and for observational errors, we allowed
the derived ages to differ by up to 5\%. Satisfactory fits with this
condition were found for all 29 of our DLEBs. Initial estimates for
$Z$ were based on the measured metallicities for the 16 systems that
have them, and for those that do not we used photometric estimates in
some cases, or even values from the literature derived from fits to
stellar evolution models, if available. We found that the final
best-fit $Z$ values occasionally deviated significantly from the
initial values, as we discuss later.

Average uncertainties in \fov, which we adopt as 0.004 for unevolved
stars (dwarfs) and 0.003 for giants, were estimated by varying this
parameter keeping \amix\ and the chemical composition fixed, with the
requirement that the predicted radii and temperatures of the binary
components be within the observational uncertainties and that the age
difference be no larger than 5\%. While these errors are considerably
more conservative than those sometimes seen in the literature, we
believe them to be realistic. The larger estimate for dwarfs reflects
the lower sensitivity to overshooting on the main sequence. A similar
procedure was followed for \amix, examining the effect of changes at
constant values of \fov\ and the chemical composition. The
uncertainties in \amix\ are estimated to be 0.20.

\section{Results}
\label{sec:results}

As a sanity check on our methods we chose the well-studied case of
Capella ($\alpha$~Aur) to generate a grid of evolutionary tracks
varying \fov\ over the range 0.000--0.025 in increments of 0.005, and
stepping \amix\ over the range from 1.7 and 1.9 every 0.1, bracketing
previous determinations for the components that used the same MESA
code \citep[see][]{Torres:15}. The observationally well constrained
$Z$ value from the same study was held fixed. In addition to verifying
that a grid search in this unambiguous case yielded essentially the
same answer as our manual method, the exercise also supports our
adopted uncertainties in \fov\ and \amix.

\begin{deluxetable*}{lcccccc}
\tablewidth{0pc}
\tabletypesize{\scriptsize}
\tablecaption{Fitted parameters for our sample of DLEBs using the GS98 mixture.\label{tab:fitted_gs98}}
\tablehead{
\colhead{} &
\multicolumn{2}{c}{Primary} & 
\multicolumn{2}{c}{Secondary} & &
\\
\colhead{Name} &
\colhead{\fov} & 
\colhead{\amix} & 
\colhead{\fov} & 
\colhead{\amix} & 
\colhead{$Z$} &
\colhead{Mean age (Myr)}
}
\startdata
SMC-108.1-14904          &  0.0210 & 1.80 & 0.0200 & 1.85 & 0.0025 &  130 \\
OGLE-LMC-ECL-CEP-0227    &  0.0180 & 2.22 & 0.0180 & 2.26 & 0.0018 &  139 \\
OGLE-LMC-ECL-06575       &  0.0190 & 2.05 & 0.0180 & 2.15 & 0.0080 &  160 \\
OGLE-LMC-ECL-CEP-2532    &  0.0170 & 2.10 & 0.0200 & 1.90 & 0.0022 &  163 \\
LMC-562.05-9009          &  0.0150 & 2.30 & 0.0150 & 2.30 & 0.0025 &  198 \\
OGLE-LMC-ECL-26122       &  0.0190 & 1.80 & 0.0170 & 2.13 & 0.0080 &  235 \\
OGLE-LMC-ECL-01866       &  0.0150 & 2.10 & 0.0150 & 2.10 & 0.0070 &  229 \\
OGLE-SMC-113.3-4007      &  0.0205 & 2.30 & 0.0205 & 2.34 & 0.0035 &  224 \\
OGLE-LMC-ECL-10567       &  0.0200 & 2.12 & 0.0180 & 2.00 & 0.0045 &  254 \\
OGLE-LMC-ECL-09144       &  0.0150 & 2.10 & 0.0180 & 1.80 & 0.0035 &  252 \\
OGLE-051019.64-685812.3  &  0.0190 & 2.40 & 0.0140 & 2.40 & 0.0045 &  275 \\
OGLE-LMC-ECL-09660       &  0.0180 & 2.06 & 0.0180 & 2.06 & 0.0035 &  341 \\
SMC-101.8-14077          &  0.0160 & 2.25 & 0.0160 & 2.28 & 0.0020 &  366 \\
$\alpha$ Aur             &  0.0200 & 1.85 & 0.0200 & 1.80 & 0.0150 &  608 \\
WX Cep                   &  0.0160 & 1.85 & 0.0190 & 1.85 & 0.0220 &  523 \\
V1031 Ori                &  0.0208 & 1.85 & 0.0190 & 1.85 & 0.0185 &  608 \\
V364 Lac                 &  0.0200 & 1.85 & 0.0200 & 1.85 & 0.0185 &  611 \\
SZ Cen                   &  0.0195 & 1.85 & 0.0190 & 1.85 & 0.0090 &  653 \\
OGLE-LMC-ECL-25658       &  0.0181 & 2.06 & 0.0181 & 2.05 & 0.0045 &  805 \\
V885 Cyg                 &  0.0190 & 1.85 & 0.0180 & 1.85 & 0.0130 &  712 \\
AI Hya                   &  0.0160 & 1.80 & 0.0180 & 1.80 & 0.0370 &  897 \\
AY Cam                   &  0.0165 & 1.80 & 0.0165 & 1.80 & 0.0175 &  970 \\
SMC-130.5-04296          &  0.0120 & 2.32 & 0.0150 & 2.04 & 0.0015 &  983 \\
OGLE-LMC-ECL-03160       &  0.0800 & 1.94 & 0.0080 & 2.15 & 0.0025 & 1023 \\
EI Cep                   &  0.0130 & 1.90 & 0.0130 & 1.90 & 0.0150 & 1309 \\
SMC-126.1-00210          &  0.0120 & 2.00 & 0.0120 & 2.05 & 0.0020 & 1240 \\
HD 187669                &  0.0090 & 1.80 & 0.0090 & 1.82 & 0.0100 & 2330 \\
OGLE-LMC-ECL-15260       &  0.0050 & 2.00 & 0.0050 & 2.11 & 0.0030 & 2143 \\
AI Phe                   &  0.0000 & 1.78 & 0.0000 & 2.05 & 0.0120 & 4383
\enddata
\end{deluxetable*}

\begin{deluxetable*}{lcccccc}
\tablewidth{0pc}
\tabletypesize{\scriptsize}
\tablecaption{Fitted parameters for our sample of DLEBs using the \as\ mixture.\label{tab:fitted_as09}}
\tablehead{
\colhead{} &
\multicolumn{2}{c}{Primary} & 
\multicolumn{2}{c}{Secondary} & &
\\
\colhead{Name} &
\colhead{\fov} & 
\colhead{\amix} & 
\colhead{\fov} & 
\colhead{\amix} & 
\colhead{$Z$} &
\colhead{Mean age (Myr)}
}
\startdata
SMC-108.1-14904         &   0.0190 & 1.80 & 0.0190 & 1.80 & 0.0018 &  123 \\
OGLE-LMC-ECL-CEP-0227   &   0.0150 & 1.95 & 0.0150 & 2.08 & 0.0022 &  138 \\
OGLE-LMC-ECL-06575      &   0.0170 & 2.02 & 0.0170 & 2.20 & 0.0070 &  155 \\
OGLE-LMC-ECL-CEP-2532   &   0.0140 & 2.00 & 0.0170 & 1.97 & 0.0021 &  155 \\
LMC-562.05-9009         &   0.0132 & 2.35 & 0.0128 & 2.35 & 0.0025 &  188 \\
OGLE-LMC-ECL-26122      &   0.0190 & 1.80 & 0.0170 & 2.13 & 0.0070 &  234 \\
OGLE-LMC-ECL-01866      &   0.0150 & 2.10 & 0.0150 & 2.10 & 0.0070 &  229 \\
OGLE-SMC-113.3-4007     &   0.0161 & 2.22 & 0.0161 & 2.22 & 0.0020 &  207 \\
OGLE-LMC-ECL-10567      &   0.0140 & 2.25 & 0.0140 & 2.00 & 0.0035 &  233 \\
OGLE-LMC-ECL-09144      &   0.0145 & 2.00 & 0.0175 & 1.75 & 0.0040 &  247 \\
OGLE-051019.64-685812.3 &   0.0160 & 2.08 & 0.0140 & 2.33 & 0.0048 &  271 \\
OGLE-LMC-ECL-09660      &   0.0170 & 2.15 & 0.0170 & 2.15 & 0.0035 &  333 \\
SMC-101.8-14077         &   0.0150 & 2.31 & 0.0150 & 2.27 & 0.0020 &  353 \\
$\alpha$ Aur            &   0.0200 & 1.83 & 0.0200 & 1.80 & 0.0120 &  594 \\
WX Cep                  &   0.0170 & 1.85 & 0.0170 & 1.85 & 0.0180 &  537 \\
V1031Ori                &   0.0180 & 1.85 & 0.0180 & 1.85 & 0.0190 &  634 \\
V364 Lac                &   0.0170 & 1.85 & 0.0170 & 1.85 & 0.0150 &  611 \\
SZ Cen                  &   0.0170 & 1.90 & 0.0160 & 1.85 & 0.0090 &  655 \\
OGLE-LMC-ECL-25658      &   0.0170 & 1.94 & 0.0170 & 1.94 & 0.0025 &  728 \\
V885 Cyg                &   0.0175 & 1.95 & 0.0170 & 1.85 & 0.0130 &  715 \\
AI Hya                  &   0.0160 & 1.87 & 0.0140 & 1.87 & 0.0220 &  956 \\
AY Cam                  &   0.0150 & 1.83 & 0.0140 & 1.83 & 0.0150 & 1023 \\
SMC-130.5-04296         &   0.0100 & 2.33 & 0.0140 & 2.04 & 0.0012 &  929 \\
OGLE-LMC-ECL-03160      &   0.0070 & 2.13 & 0.0070 & 2.13 & 0.0033 & 1034 \\
EI Cep                  &   0.0110 & 1.85 & 0.0110 & 1.80 & 0.0130 & 1299 \\
SMC-126.1-00210         &   0.0100 & 1.95 & 0.0100 & 1.95 & 0.0013 & 1156 \\
HD 187669               &   0.0090 & 1.80 & 0.0090 & 1.80 & 0.0075 & 2204 \\
OGLE-LMC-ECL-15260      &   0.0040 & 2.00 & 0.0040 & 2.25 & 0.0040 & 2197 \\
AI Phe                  &   0.0000 & 1.70 & 0.0000 & 1.95 & 0.0090 & 4108
\enddata
\end{deluxetable*}

\begin{figure*}
  \epsscale{1.15}
  \plotone{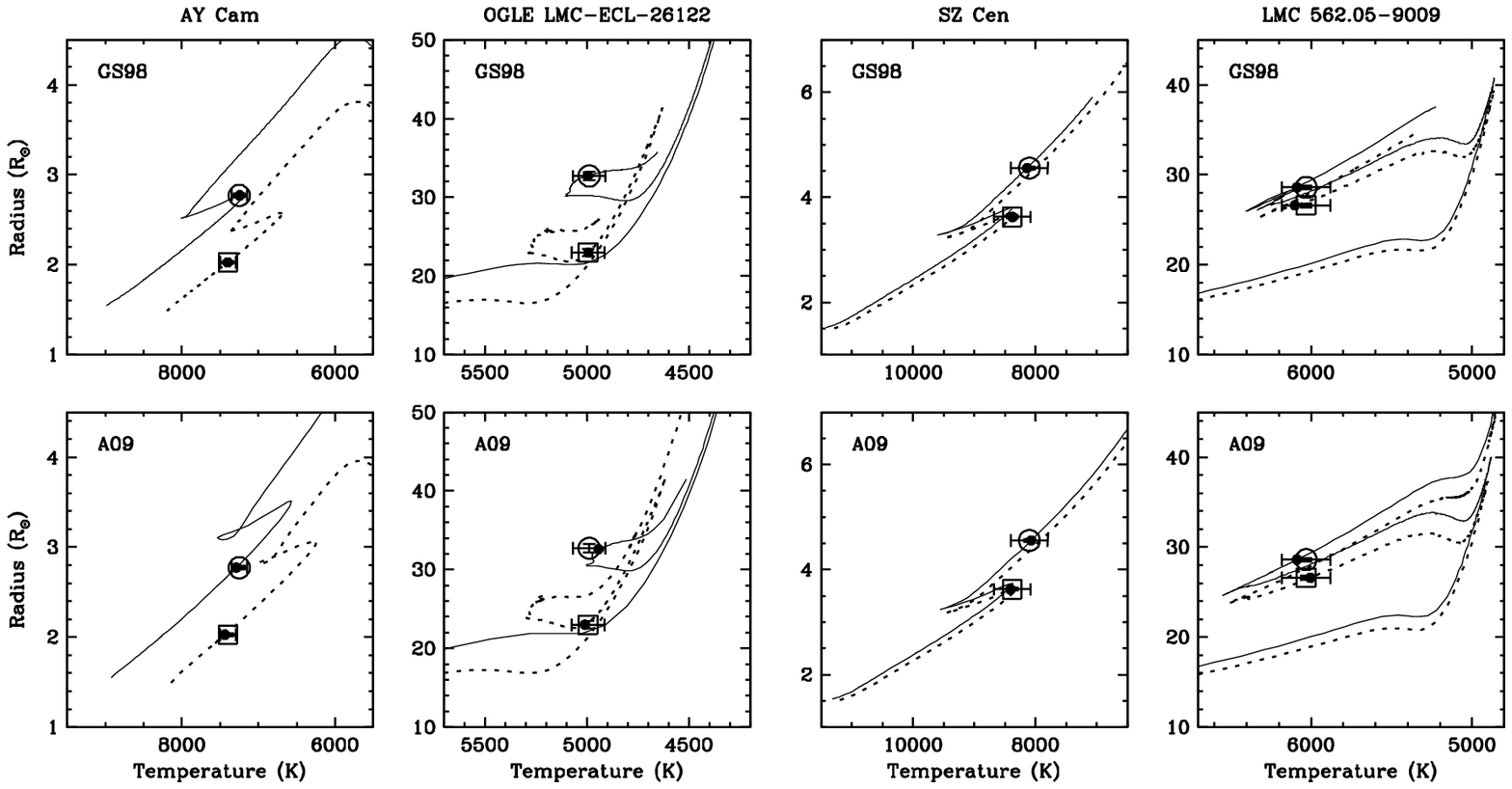}
  \figcaption{Sample best fits to four of our binaries in the $R$
    vs.\ $T_{\rm eff}$ diagram, for the element mixtures of \gs\ (top
    row) and \as\ (bottom row). Evolutionary tracks and the
    observations for the primary in each system are represented with
    solid lines and open circles, while dotted lines and open squares
    are used for the secondary. Small dots mark the best-fit location
    on each track, and are always within the measurement
    uncertainties. \label{fig:rteff} }
\end{figure*}

The results of our calculations for the \gs\ mixture are presented in
Table~\ref{tab:fitted_gs98} with the systems listed in the same order
as in Table~\ref{tab:sample}, and include the inferred values of the
overshooting and mixing length parameters (\fov, \amix) for each
component, as well as the optimal abundance $Z$ and mean age for each
system.  Table~\ref{tab:fitted_as09} contains the corresponding
results for the \as\ mixture. Representative fits for four of our
systems are shown in Fig.~\ref{fig:rteff}, with the top row
corresponding to the \gs\ set and the bottom row to \as. The
morphology of the evolutionary tracks as well as the inferred
evolutionary state of the components is quite similar for the two
mixtures.

\begin{figure*}
  \epsscale{1.10}
  \plotone{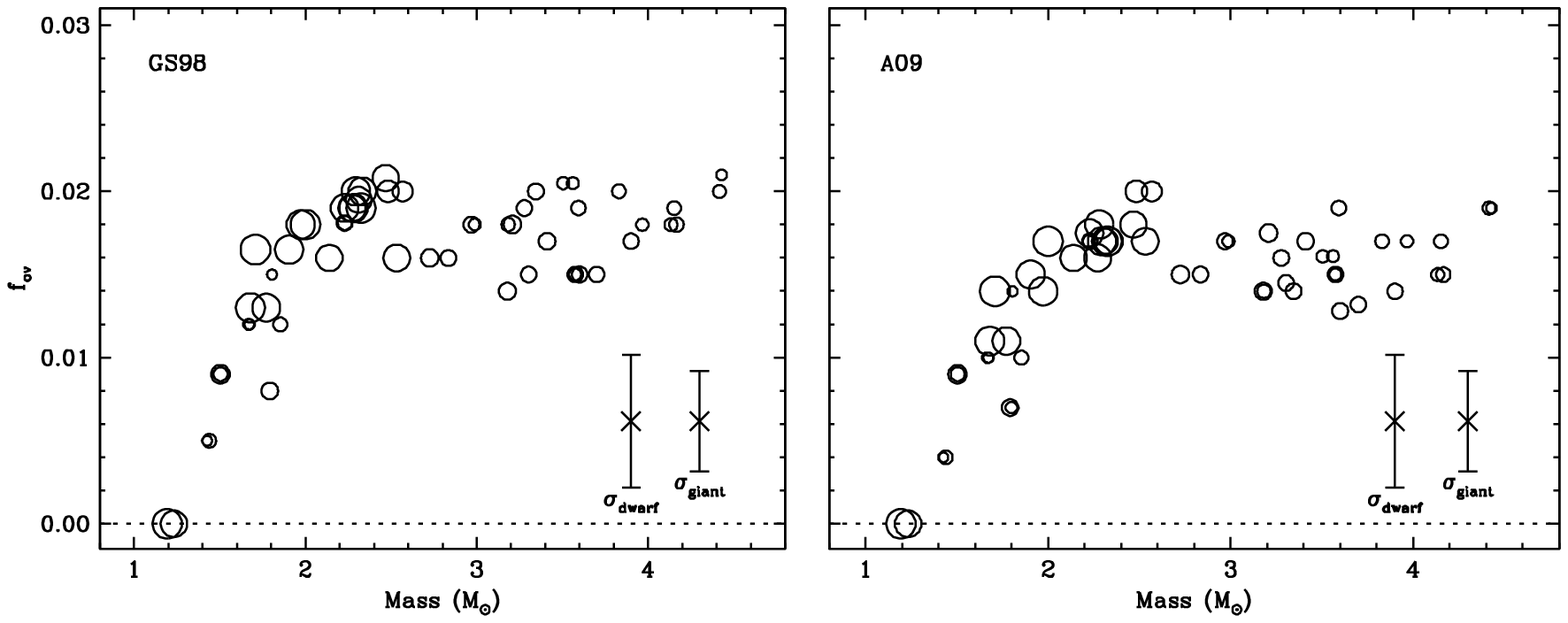}
  \figcaption{Semi-empirical determinations of the overshooting parameter
    \fov\ for both element mixtures, shown as a function stellar mass
    for the 29 systems in our sample. The size of the points is
    proportional to $\log g$. Typical uncertainties for dwarfs and
    giants are represented with the error bars on the lower
    right. \label{fig:over}}
\end{figure*}

The behavior of \fov\ for each element mixture as a function of
stellar mass is displayed in Fig.~\ref{fig:over}, in which the size of
the symbols is proportional to the surface gravity $\log g$. All
systems more massive than about $2.5~M_{\sun}$ are giants.
To
avoid clutter we show the mean error bars for evolved and unevolved
systems on the lower right, rather than on each point.

We find a clear dependence of the overshooting parameter on mass for
both mixtures, with a strong rise in \fov\ up to about $2~M_{\odot}$
that flattens out thereafter. This represents the first semi-empirical
mass calibration of \fov. We defer a discussion of the nature of this
relation until Section~\ref{sec:coresize}.  The trend appears
qualitatively similar for \gs\ and \as, though there may be slight
differences. For example, if we consider the regime of stars more
massive than $2~M_{\odot}$, where \fov\ is effectively constant, we
find an average overshooting parameter for the sample of $\fov =
0.0181 \pm 0.0003$ for the \gs\ set (mean of estimates for 40 stars)
and $\fov = 0.0164 \pm 0.0003$ for the \as\ set.\footnote{This later
  value happens to be similar to the one adopted in the MIST grid
  \citep{Choi:16} for {\it all} stars, regardless of their mass.}
While these appear statistically distinct, the formal uncertainties do
not include possible systematic errors, which are difficult to
quantify and may be much larger than the statistical errors, so we
hesitate to conclude there is a real difference.

Concerning our fitted \amix\ values, which rely on the 1-D MESA code,
we find that they are in reasonable agreement with the predictions
from 3-D radiative hydrodynamic simulations by \cite{Magic:2015}
restricted to the same range of metallicities, effective temperatures,
and surface gravities as the binaries in our sample. This is
noteworthy considering the significant differences between the input
physics of the MESA code and the 3-D models of Magic et al., and the
presence of observational errors in our own estimates.

We note also that the model comparisons with the \gs\ mixture lead to
best-fit $Z$ values that are on average slightly but systematically
smaller than the measured abundances, for the 16 systems for which
empirical metallicity estimates are available. A similar but more
pronounced effect was seen in Paper~I, and is most obvious for some of
the metal-poor systems in the LMC and SMC. The average disagreement is at about
the 2$\sigma$ level, corresponding to a [Fe/H] difference of $0.11 \pm
0.05$~dex. We see no such discrepancy on average with the \as\ mixture
($\Delta{\rm [Fe/H]} = 0.02 \pm 0.05$~dex). Possible explanations for
the difference in behavior may be related to the opacities themselves,
an incorrect slope $\Delta Y/\Delta Z$, or an inaccurate value for the
adopted primordial helium content for the \gs\ mixture.  We return to
this in the next section.

\section{Discussion}
\label{sec:discussion}

\subsection{The diffusive and step-function implementations of
overshooting compared}
\label{sec:comparison}

A first interesting conclusion we draw from the results is that the
qualitative way in which the strength of the overshooting depends on
stellar mass (Fig.~\ref{fig:over}) seems essentially independent of
the opacities and helium enrichment laws involved. Furthermore, the
run of \fov\ with mass from both the \gs\ and \as\ sets is in turn
qualitatively very similar to that obtained in Paper~I for \aov, which
used the step-function approximation. Thus, the details of how the
phenomenon is parametrized also seem unimportant, at least when it
comes to the mass dependence. While this last conclusion might have
been anticipated from the rough similarity between the internal
stellar structures inferred theoretically using the two overshooting
formulations \citep[e.g.,][]{Magic:10}, empirical verification such as
we report here remains essential, can lead to further insights, and
now provides a practical recipe for incorporating the \fov\ mass
dependence into models, in the same way as our earlier work in Paper~I
supplied one for the classical \aov\ implementation. We note, however,
that with our current understanding of the phenomenon one recipe
cannot necessarily be inferred from the other.

A number of authors \citep[e.g.,][]{Herwig:97, Noels:10} have in fact
suggested an approximate scaling between the two overshooting
parameters such that $\aov/\fov \approx 10$, though very few
semi-empirical estimates of this relation seem to be available in the
literature.  One estimate of $\aov/\fov \approx 13$ was reported in an
asteroseismic study of the (single) slowly pulsating B star
KIC~7760680 by \cite{Mora:16}. A study of the eclipsing binary TZ~For
by \cite{Valle:17} gave $\aov/\fov \approx 12$.  \cite{Magic:10} made
a direct comparison between models for masses ranging from
2~$M_{\sun}$ to 6~$M_{\sun}$ using the step-function approximation
(with \aov\ held fixed at 0.20) and calculations based on the
diffusive implementation, and obtained a best-fit value with the
latter of $\fov = 0.018$, corresponding to a ratio $\aov/\fov \approx
11$. However, their study did not address the more critical regime
below 2~$M_{\sun}$, or stars that have evolved off the main sequence.

\begin{figure}
  \epsscale{1.10}
  \plotone{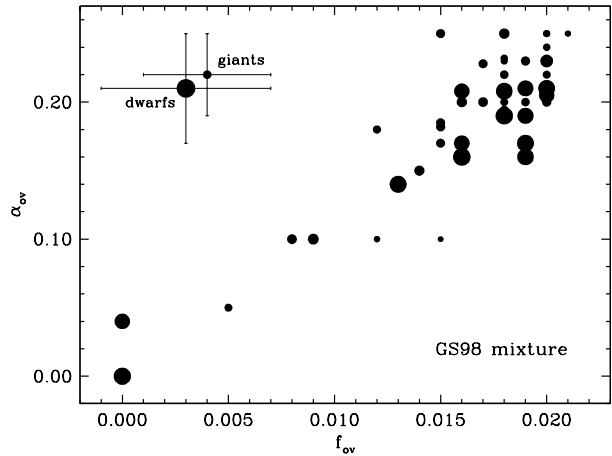}

  \figcaption{Values of \aov\ from Paper~I shown against the
      \fov\ values from this paper. Both use the GS98 mixture. The
      symbol size is proportional to $\log g$, as in
      Fig.~\ref{fig:over}, and representative error bars for evolved
      and unevolved stars are shown on the upper left.
    \label{fig:aov_fov} }
\end{figure}

A star-by-star comparison between the \fov\ values from the present
work and the \aov\ values from Paper~I (both using the \gs\ mixture
and the same enrichment law) enables us to revisit this issue. We find
that on average $\aov/\fov = 11.36 \pm 0.22$ (mean of 56 ratios,
excluding the two stars in AI~Phe with $\fov = 0$). This is consistent
with previous estimates, but is based on a much larger sample of
semi-empirical determinations.  The two parameters plotted against
each other are shown graphically in Fig.~\ref{fig:aov_fov}.
Closer inspection suggests, however, that the connection
may be more complex, and there may be a slight dependence of the
$\aov/\fov$ ratio on the surface gravity $\log g$, or possibly $Z$,
mass, or $T_{\rm eff}$. Unfortunately these quantities are strongly
correlated with each other in our sample. This is because the majority
of the cool giants are in the LMC and SMC and are both more massive
and significantly more metal-poor than the field dwarfs, so it is not
possible to ascertain which variable is the most relevant one.
Nonetheless, as an illustration, if we split the sample at $T_{\rm
  eff} = 6500$~K we obtain averages of $\aov/\fov = 10.50 \pm 0.25$
for the hotter stars and $\aov/\fov = 11.71 \pm 0.27$ for the cooler
ones. Identical numbers result for the dwarfs and giants,
respectively, if we split the sample at $\log g = 3.0$. The difference
is formally at the 3$\sigma$ level. An even larger sample with a range
of uncorrelated stellar properties would be needed to investigate this
issue further.

Overshooting has a significant impact on stellar ages.  Between the
present work and that of Paper~I we now have in hand semi-empirical
estimates of the overshooting parameter as well as age estimates
derived in the framework of two separate approaches (\fov\ and \aov)
for 29 binary systems, using the same element mixture (\gs) and the
same helium enrichment law.  We find that the mean ages for these 29
systems in common are rather similar for the two overshooting
implementations, with an average ratio of ${\rm Age(\fov)/Age(\aov)} =
0.975 \pm 0.002$ (statistical error) and a range of values between
0.88 for AI~Phe and 1.10 for OGLE~LMC-ECL-26122.  Although the mean
ratio is formally smaller than unity, we suspect the difference may
not be significant given that unquantified systematics probably
dominate the error budget, and especially that there are also
differences between the Granada and MESA evolutionary codes used in
each case, both in terms of their physical ingredients and their
numerical details.

\subsection{The size of the convective core and the nature of the mass dependence of \fov}
\label{sec:coresize}

\begin{figure}
\epsscale{1.10}
\plotone{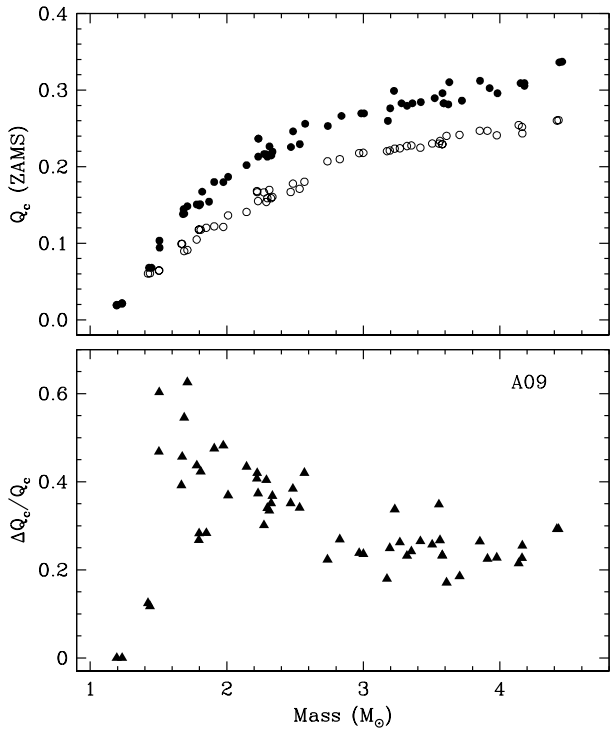}

\figcaption{{\it Top:} Values of the mass $Q_c$ of the convective core
  (normalized to the total stellar mass) extracted from the best-fit
  evolutionary track for each star in our sample at the point
  corresponding to the ZAMS (filled symbols).  The results shown are
  for the \as\ mixture; those for the \gs\ mixture are similar. The
  open symbols represent $Q_c$ also at the ZAMS for standard models
  with no overshooting, at the same chemical composition. The
  difference in core size between the filled and open symbols that is
  due to overshooting is seen to vary with stellar mass. {\it Bottom:}
  Fractional increase in $Q_c$ between the filled and open circles in
  the top panel.} \label{fig:qc}

\end{figure}

In this section we comment on the underlying reasons for the initial
increase in \fov\ with mass shown in Figure~\ref{fig:over}, starting
around $M \approx 1.2~M_{\sun}$, and the subsequent flattening of the
relation after about 2~$M_{\sun}$ and up to the mass limit of our
sample. For this we focus on the behavior of the size of the
convective core, as indicated by the models, which is a quantity more
closely linked to the interior physics.  Using our best-fit values for
\fov, \amix, and the chemical composition for each star, we have
extracted from the corresponding evolutionary tracks the mass of the
convective core $Q_c$ that the star has at the zero-age main sequence
(ZAMS)\footnote{The ZAMS in our models is defined as the location on
  the H-R diagram at which the central hydrogen content drops to
  99.4\% of its initial value.}, which is directly determined by the
strength of the overshooting as parametrized by \fov.  We restrict
ourselves to examining the physics at the ZAMS in order to avoid
effects from the evolution that occurs later, and because of the
simpler, nearly homogeneous chemical structure of ZAMS stars.  The top
panel of Fig.~\ref{fig:qc} shows how this quantity $Q_c$, normalized
to the total mass of the star, varies as a function of mass for the
\as\ element mixture in the diffusive approach to overshooting used in
this paper (filled symbols). We find a very similar trend for the
\gs\ mixture (not shown), indicating the opacities have little
influence.

The open symbols in the top panel of Fig.~\ref{fig:qc} represent the
normalized mass of the convective core at the ZAMS {\it in the
  absence of overshooting} (i.e., from ``standard'' models that use
the Schwarzschild criterion to set the core boundary). As before, the
results are similar for \gs\ and \as. The separation between the
filled and open symbols at a given mass represents the enlargement of
the convective core due to overshooting. This net growth is seen to
increase rapidly with mass up to about $2~M_{\sun}$, but not much
thereafter, reflecting the same trend observed in
Fig.~\ref{fig:over}. An equivalent representation is displayed in the
lower panel of Fig.~\ref{fig:qc}, which shows the fractional change in
$Q_c$. For both element mixtures we see an obvious trend such that the
fractional increase in $Q_c$ is larger for the less massive stars,
reaching $\sim$60\% at about 1.5~$M_{\sun}$ and then decreasing to
20--30\% for the higher mass stars in our sample.

The physical reasons for this behavior of the fractional size of the
convective core as a function of stellar mass, with a change in
character around $2~M_{\sun}$, have to do with concomitant changes in
the opacities, the equation of state, and the nuclear reaction
rates. We explore this in further detail in the Appendix using a
simple analytical toy model involving homology transformations and the
basic differential equations of stellar structure. We show there that
it is indeed possible to understand the trends in
Figures~\ref{fig:over} and \ref{fig:qc} as due at least in part to the
role of those ingredients at different masses.

Finally, we note that the way in which $Q_c$ varies with stellar mass
is essentially the same as seen in Figure~5 of Paper~I for the
step-function approximation to overshooting, and similarly with the
fractional change in $Q_c$ (Figure~5 of Paper~I, lower panel). In the
latter case, though, the effect was somewhat magnified because the
models without overshooting were computed for a single representative
solar metallicity, rather than the $Z$ value of each system that we
use in the present work.  Thus, once again the formulation of
overshooting appears not to matter much in terms of the general trend,
at least to first order. This is not entirely surprising given the
approximate scaling reported earlier between the two free parameters,
$\aov \approx 11.4 \fov$, which to some extent implies a
correspondence in the net effect of overshooting even though the
details of how it operates in the two implementations are not the
same.

\subsection{Relation between the best-fit metallicities and the primordial
helium abundance}

In Section~\ref{sec:results} we noted that our fitting procedure with
the \gs\ mixture resulted in a small but systematic difference, on
average, between the best-fit abundances and those measured for the 16
binary systems where this is available. The fitted values are slightly
smaller, especially for the more metal-poor systems. This is in the
same direction as a similar discrepancy noted in Paper~I, which
explored the classical prescription for overshooting with the same
\gs\ element mixture and also the same helium enrichment law used here
($Y_p = 0.24$, $\Delta Y/\Delta Z = 2.0$). It was pointed out in our
earlier work that the effect could possibly be explained if the
measured temperatures for all the binaries were systematically too hot
by 150--200~K, though this seemed unlikely.

Interestingly, the abundance discrepancy is not present when using the
\as\ mixture and an alternate enrichment law specified by $Y_p =
0.249$ and $\Delta Y/\Delta Z = 1.67$. To investigate which, if any,
of these two ingredients (mixture, or enrichment law) might be
responsible for the metallicity disagreements, we repeated the fits
for some of the most obvious outliers that have a measured
metallicity, using again the \gs\ mixture but this time with the above
enrichment law instead of the one indicated in the previous paragraph.
We obtained virtually unchanged values for \fov, \amix, and the mean
system ages, and the resulting best-fit metallicities were now in
better agreement with the observations, pointing to the helium
abundance as the culprit. More specifically, the improvement is
largely the result of the increase in $Y_p$ from 0.24 to 0.249, rather
than the change in the slope $\Delta Y/\Delta Z$ from 2.0 to 1.67,
because the slope difference has only a minor impact on the total
helium abundance for our metal-poor binary systems, which are the ones
where the metallicity discrepancy is most noticeable.  Similar results
were obtained from experiments with the Granada code and the
step-function implementation of overshooting using $Y_p = 0.249$,
again suggesting a connection with helium. In support of these clues
we note that both the radius and temperature of a given stellar
configuration depend on some positive power of the mean molecular
weight. A reduction in the hydrogen content of the core implies a
larger mean molecular weight, which in turn increases the radius and
temperature at a given mass.  Thus, at a fixed value of $Z$, models
with the more recent determination of $Y_p = 0.249$ \citep{Planck:16}
are hotter, larger, and brighter than those with $Y_p = 0.24$, which
explains at least in part why the use of a lower $Y_p$ with the
\gs\ mixture seemed to imply the empirically measured temperatures
were overestimated. We point out, finally, that this realization does
not affect the conclusions by \cite{CT:16} regarding the mass
dependence of \aov, or the similar conclusions reported here with the
diffusive overshooting formulation and the \gs\ mixture. This is
because neither the \aov\ nor the \fov\ inferences are significantly
influenced by the assumed helium abundance as long as $Z$ is allowed
to vary in the fits, as has been the case both here and in Paper~I.

\section{Concluding remarks}
\label{sec:conclusions}

The key result of this paper is the determination of the explicit
dependence of the diffusive convective core overshooting parameter
\fov\ on stellar mass in a semi-empirical way, relying on a set of
nearly 30 eclipsing binary systems with well measured properties. This
study extends our previous one of \cite{CT:16} that used the classical
step-function approximation for overshooting with its free parameter
\aov\ and a similar set of binaries, and completes our investigation
of the phenomenon with the two most commonly used prescriptions.
Additionally, in the present work we have tested two different element
mixtures for the opacities, adding to the \cite{Grevesse:98} mixture
used in Paper~I a more recent mixture by \cite{Asplund:09}.  We find a
clear dependence of \fov\ on stellar mass rising to about 2~$M_{\sun}$
and then leveling off to the upper mass limit of our sample
($\sim$4.4~$M_{\sun}$), regardless of the opacities. This qualitative
behavior is similar to the one found before for \aov. While many
recent grids of stellar evolution models build in variations in the
strength of overshooting with mass in somewhat arbitrary ways, this
study and that in Paper~I now provide practical recipes for doing this
that are grounded on observations, rather than expectations from
theory. Our \fov\ calibration therefore represents an important step
forward, although there is still ample room for improvement before it
can be considered definitive.  For example, our binary sample has few
stars near the low end of the mass range where overshooting begins to
ramp up, and about half of the systems are lacking empirical
metallicity estimates that could be used to further constrain the
fits.  Observational errors and shortcomings in the models themselves
must also be kept in mind.

We have examined the nature of the mass dependence of \fov\ by
focusing on the related fractional increase in the size of the
convective core at the ZAMS ($\Delta Q_c/Q_c$) caused by diffusive
overshooting. This quantity is more closely linked to the physics of
stellar interiors, and also shows a mass dependence such that stars
with smaller masses have larger relative increases in the core mass
$Q_c$ up to 60\%, which becomes smaller (20--30\%) for larger
stars. The behavior is similar for both element mixtures used here,
and also mirrors that seen in Paper~I with the step-function
implementation of overshooting. The use of a simple model based on
homology transformations and the basic differential equations of
stellar structure succeeds in explaining this trend in a qualitative
way, revealing the role of opacities, the equation of state, and
nuclear reaction rates.  A deeper understanding of the behavior of
\fov\ as a function of stellar mass displayed in Figure~\ref{fig:over}
would benefit from an investigation into the physical conditions at
the edge of the convective cores, and how they affect the convective
plumes. While such a study is beyond the scope of the present work, we
plan to undertake it in a future publication.

Our use of two different helium enrichment laws associated with the
adopted \gs\ and \as\ mixtures has also helped to shed light on a
nagging systematic difference between the fitted and measured
metallicities noted originally in Paper~I, and seen also here with the
\gs\ set to a lesser degree. We find that the discrepancy is largely
removed by adopting the higher primordial helium abundance $Y_p =
0.249$ established in recent studies \citep{Planck:16}, instead of the
value $Y_p = 0.24$ we used previously.

Comparing our \fov\ determinations that use the \gs\ mixture and the
diffusive approximation with the \aov\ estimates from Paper~I based on
the step-function prescription and the same mixture (as well as the
same enrichment law), we establish a semi-empirical scaling
relationship between the two overshooting parameters such that
$\aov/\fov = 11.36 \pm 0.22$ (mean of 56 estimates), with a tentative
dependence on either temperature or $\log g$, or possibly mass or $Z$
(all strongly correlated in our sample). This connection between
\fov\ and \aov\ is consistent with, but now more firmly established
than previous estimates.

\begin{acknowledgements}
We are grateful to A.\ Dotter for expert assistance and advice in
using the MESA module, as well as C.\ Aerts for references and helpful
comments on the subject of this paper. We also thank our two
  anonymous referees for constructive criticism. The Spanish MEC
(AYA2015-71718-R and ESP2015-65712-C5-5-R) is gratefully acknowledged
for its support during the development of this work. GT acknowledges
partial support from the NSF through grant AST-1509375. This research
has made use of the SIMBAD database, operated at the CDS, Strasbourg,
France, and of NASA's Astrophysics Data System Abstract Service.
\end{acknowledgements}

\begin{appendix}
\label{sec:appendix}

\section{Core overshooting and homology transformations}

Here we explore and attempt to understand the nature of the stellar
mass dependence of $\Delta Q_c/Q_c$ (the relative change in the core
mass shown in the lower panel of Fig.~\ref{fig:qc}), which is directly
linked to the \fov\ trend of Fig.~\ref{fig:over}, as revealed by our
semi-empirical estimates for our sample of 29 eclipsing binary
systems.  We will use simple ideas based on homology transformations,
in which the two stellar models to be scaled are required to have the
same relative mass distribution. One of the most useful tools to
investigate how mass is distributed in stellar interiors is the Radau
differential equation, whose solution provides the apsidal-motion
constant $k_j$ of order $j$ that gives an accurate indication of the
mass distribution. Standard models and those computed with core
overshooting for the same mass present similar apsidal-motion
constants at the ZAMS, indicating that they also have similar relative
mass distributions \citep[see, e.g.,][Figs.~2, 4, and
  6]{Claret:91}. The same holds when analyzing the properties of the
respective cores. Therefore, we may apply homology transformations to
both standard homogeneous models as well as those with extra mixing
from overshooting. Due to the complexity of the various ingredients of
stellar physics (opacities, energy generation, equation of state,
etc.) homology transformations do not guarantee accurate solutions,
but they are still very useful to explore certain problems in a
qualitative way, and even quantitatively to some degree, providing
insights into the physics involved.

For the problem at hand ---understanding the change in $\Delta
Q_c/Q_c$ as a function of stellar mass, which is closely related to
the change in \fov\ as a function of mass--- we begin with a star in
hydrostatic equilibrium with a convective core of mass $Q_0$ that
satisfies the Schwarzschild criterion, and that is surrounded by a
shell of negligible mass in radiative equilibrium.  Consider now a
second model to be scaled from the first, with a large convective core
$Q_0 + \Delta Q$. Our reference shell is displaced upward to a
position $R_0 + \Delta R$ simulating the effect of core
overshooting. We apply simple homology relations and the basic
differential equations of stellar structure and evolution to study in
an approximate quantitative way the impact of the increase in radius
(and mass) of the convective core.

The net rate of nuclear energy generation inside a sphere of radius
$r$ is given by
\begin{eqnarray}
 \frac{dL_g(r)}{dr} =  4 \pi r^2\rho\epsilon,
\end{eqnarray}
where $L_g$ denotes the rate of energy production by nuclear
processes, $\epsilon$ is the thermonuclear energy generation rate per
unit mass ($\epsilon = b\rho T^{\nu}$), $\rho$ is the density, $T$ is
the temperature, $b$ is a constant, and $r$ is the radial distance.
Here we focus on homogeneous (ZAMS) models to compare with the data
displayed in Fig.~\ref{fig:qc}. Let $Q_0$, $R_0$ and $\mu_0$ be the
mass, radius, and the mean molecular weight of the standard convective
core (without extra mixing) and $Q$, $R$, and $\mu$ the same variables
for a configuration with extra mixing, i.e., with radius $R= R_0 +
\Delta R$, core mass $Q= Q_0 + \Delta Q$, and mean molecular weight
$\mu= \mu_0 + \Delta \mu$. The homology transformations will be used
to scale one model to the other and evaluate the changes $\Delta Q$
and $\Delta R$ produced by the additional mixing.  Inserting the usual
homology transformations for the density and temperature, assuming an
ideal non-degenerate gas with no radiation pressure, and integrating
the above equation, we obtain
\begin{eqnarray}
L_g(x) = {\frac{Q}{Q_0}}\int_{0}^{x}\left({\frac{b}{b_0}}\right)_{x'}\frac{dL_{go}(x')}{dx'}dx' = 
L_{go}(x)\left({\frac{b}{b_0}}\right)\left({\frac{\mu}{\mu_0}}\right)^{\nu}
\left({\frac{Q}{Q_0}}\right)^{2+\nu} \left({\frac{R}{R_0}}\right)^{-3 -\nu}
\end{eqnarray}
where $x=r/R$.

The restriction mentioned above concerning the equation of state is
necessary because, even if we consider an ideal gas, the addition of
radiation pressure will not lead to permanent homology relationships.
On the other hand, as we are dealing with homogeneous models, we may
take $\mu = \mu_0$, and we may also assume that the constants $b$ and
$b_0$ are independent of $x$ for small $\Delta R$.  In this case, the
relative change in the nuclear energy generation will be given by
\begin{eqnarray}
 \frac{\Delta L_g}{L_g} =  (2 + \nu)\frac{\Delta Q}{Q_0} - (3+\nu)\frac{\Delta R}{R_0}.
\end{eqnarray}
Additionally, the equation of radiative transfer can be written as 
\begin{eqnarray}
 \frac{\partial T}{\partial r}   =  - \frac{3}{16\pi ac}\frac{\kappa\rho L_{\rm rad}}{r^2 T^3}, 
\end{eqnarray}
where $L_{\rm rad}$ is the net rate of energy carried out by
radiation, $a$ is the radiation pressure constant, and $c$ is the
speed of light.  As before, we apply the homology relation for the
density assuming a perfect ideal gas, no radiation pressure, and
assuming a Kramers complete opacity law given by $\kappa = B \rho^n
T^{-s}$, where $B$ is a constant.  The same remarks from above
concerning $b$ and $b_0$ hold also for the constants $B$ and
$B_0$. Finally the resulting relative change of energy $L_{\rm
  rad}$ is
\begin{eqnarray}
 \frac{\Delta L_{\rm rad}}{L_{\rm rad}} =  (3 -n + s)\frac{\Delta Q}{Q_0} - (3n-s)\frac{\Delta R}{R_0}.
\end{eqnarray}

We note that the relative change in $L_{\rm rad}$ is insensitive to
the rate of nuclear energy generation because it is independent of
$\nu$.  Under conditions of thermal equilibrium we have $\Delta
L_g/L_g = \Delta L_{\rm rad}/L_{\rm rad}$, from which the fractional
increase in the convective core mass due to the extra mixing is easily
derived as
\begin{eqnarray}
\label{eq:dqq}
\frac{\Delta Q}{Q_0}= \frac{(3+3n+\nu-s)}{(n+\nu-s-1)}\frac{\Delta R}{R_0}.
\end{eqnarray}

Note also that $L_{\rm rad}$ depends only on how energy is transported
(opacities, equation of state), whereas $L_g$ depends on how it is
generated (thermonuclear energy generation rate per unit mass).  The
functional form of the slope at each mass in Eq.~\ref{eq:dqq} is the
result of the interaction between both processes.  This interaction is
a consequence of the imposition of the condition of thermal
equilibrium.  The above equation may also be derived in a different
way using permanently homologous models by expanding $Q/Q_0$ in a
series up to 2nd order.

The slope $(3+3n+\nu-s)/(n+\nu-s-1)$ in Eq.~\ref{eq:dqq} presents
some interesting features.  For pure Thomson scattering or for Kramers
opacities it is larger for lower values of $\nu$, that is, for stars
for which the proton-proton chain contributes significantly to the
energy generation. In other words, the slope is more pronounced for
less massive stars than for more massive ones.  For larger values of
$\nu$ the slope has a near-asymptotic behavior, independently of the
adopted opacity law.

The dependence on temperature of the two major nuclear sources of
energy during the main-sequence (the proton-proton chain and the CNO
cycle) shows three main characteristics: (1) for $\log T < 7.25$ the
dominant source is the proton-proton chain, (2) for $\log T \approx
7.25$ the two processes have a similar contribution, and (3) for $\log
T \approx 7.3$ (corresponding to a stellar mass larger than about
2~$M_{\sun}$) the CNO cycle contributes about 10 times more than the
proton-proton chain, and at a slightly higher temperature of $\log T =
7.4$ it becomes about 100 times more important. Due to the complicated
nature of the nuclear energy rates, and the fact that in a given star
both nuclear processes may be acting at the same time in different
regions, it is difficult to determine an effective value of $\nu$ for
each mass.  It is common to assign a typical value of $\nu = 4$ for
stars of small mass and $\nu = 17$ for more massive stars, though
without explicitly stating a dependence on mass. To provide a
continuous relation between these two regimes for our purposes we
adopt a simple linear expression motivated by the properties of the
nuclear energy production summarized above, of the form $\nu_{\rm eff}
= (M-1.25) + 4.8$, to be inserted in Eq.~\ref{eq:dqq}.  In this
expression $M$ is the stellar mass in solar units. For simplicity we
assume also that the opacity is given by the classical Kramers law ($n
= 1$, $s=7/2$).

\begin{figure}
  \epsscale{0.60}
  \plotone{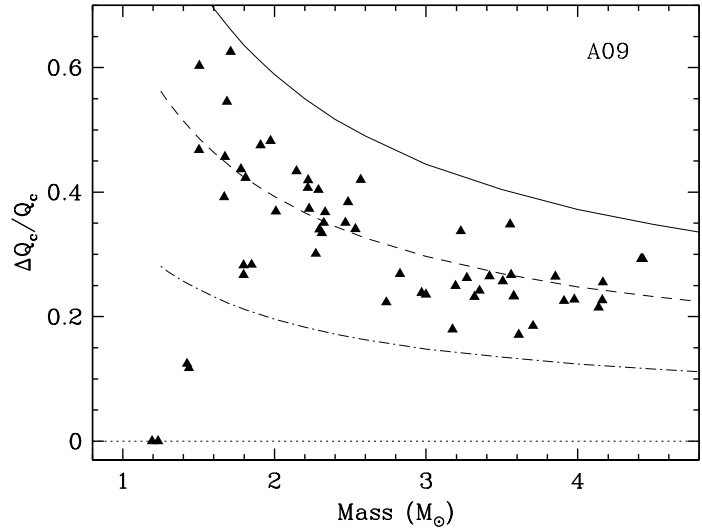}
  \figcaption{Relative increase in the mass of the convective core as a
    function of stellar mass.  The solid line corresponds to the
    iso-$\Delta R/R_0$ curve for the value 0.15, the dashed line is
    for 0.10, and the dot-dashed line for 0.05. Triangles represent the
    semi-empirical values with the \as\ mixture. \label{fig:dqq}}
\end{figure}

With these ingredients we may now compare the relative increase in
core mass due extra mixing, as given by Eq.~\ref{eq:dqq}, with the
semi-empirical data from our 29 DLEBs. As we cannot establish the
values of $\Delta R/R_0$ for the binary components a priori (from
first principles), we consider a few fixed values and draw iso-$\Delta
R/R_0$ curves in Fig.~\ref{fig:dqq} for comparison with the
observations. The triangles in this figure represent the
semi-empirical values of $\Delta Q_c/Q_c$ for each star obtained using
the \as\ mixture. We explore three values of $\Delta R/R_0$ as shown
by the three curves in the figure (0.15, 0.10, and 0.05, from top to
bottom). Detailed estimates of $\Delta R/R_0$ for selected binaries in
our sample using the best-fit models give values within this range,
indicating they are realistic.

Despite the simplicity of our analytical model, the theoretical
iso-$\Delta R/R_0$ curves show a similar pattern of variation as the
semi-empirical data, with the less massive stars displaying higher
$\Delta Q_c/Q_c$ values than more massive stars. Additionally, it can
be seen that the previously mentioned near-asymptotic behavior of the
slope for the more massive stars is consistent with the semi-empirical
results. While these calculations were made specifically for the
\as\ mixture, we found the same features when using the \gs\ set.

In summary, building on very simple approximations Eq.~\ref{eq:dqq}
is able to successfully predict the pattern of changes in the relative
increase $\Delta Q_c/Q_c$ in the mass of the convective core due to
overshooting, illustrating the connection between this slope and
stellar mass, the type of nuclear process, the opacities, and the
equation of state.

\end{appendix}


\begin{thebibliography}

\bibitem[Planck Collaboration(2016)]{Planck:16} Ade, P.A.R., Aghanim,
  N., Arnaud, M.  et al. 2016, \aap, 594, A13

\bibitem[Aerts(2013)]{Aerts:13} Aerts, C. 2013, in Setting a New
  Standard in the Analysis of Binary Stars, eds.\ K.\ Pavlovski,
  A.\ Tkachenko \& G.\ Torres, EAS Publications Series, Vol.\ 64,
  2013, pp.\ 323-330

\bibitem[Asplund et al.(2009)]{Asplund:09} Asplund, M., Grevesse, N.,
  Sauval, A.\ J., \& Scott, P. 2009, \araa, 47, 481 (A09)

\bibitem[B\"ohm-Vitense(1958)]{Vitense:58} B\"ohm-Vitense, E. 1958,
  \zap, 46, 108

\bibitem[Choi et al.(2016)]{Choi:16} Choi, J., Dotter, A.,  Conroy, 
C. et al 2016, \apj, 823, 102C

\bibitem[Claret \& Gim\'enez(1991)]{Claret:91} Claret, A., Gim\'enez,
A. 1991, \aap, 244, 319

\bibitem[Claret(2004)]{Claret:04} Claret, A. 2004, \aap, 424, 919

\bibitem[Claret(2007)]{Claret:07} Claret, A. 2007, \aap, 475, 1019

\bibitem[Claret \& Gim\'enez(2010)]{Claret:10} Claret, A., Gim\'enez,
A. 2010, \aap, 519, 57

\bibitem[Claret \& Torres(2016)]{CT:16} Claret, A., Torres, G.  2016, 
\aap, 592, A15 (Paper~I)

\bibitem[Deheuvels et al.(2016)]{Deheuvels:16} Deheuvels, S., Brandao, 
I., Silva Aguirre, V., et al. 2016, \aap, 589A, 93D.

\bibitem[Elgueta et al.(2016)]{Elgueta:16} Elgueta, S.\ S., Graczyk,
  D., Gieren, W.\ et al.\ 2016, \aj, in press

\bibitem[Fekel et al.(2013)]{Fekel:13} Fekel, F.\ C., Henry, G.\ W.,
  \& Sowell, J.\ R. 2013, \aj, 146, 146

\bibitem[Ferguson et al.(2005)]{Ferguson:05} Ferguson, J.\ W.,
  Alexander, D.\ R., Allard, F., Barman, T., Bodnarik, J.\ G.,
  Hauschildt, P.\ H., Heffner-Wong, A., \& Tamanai, A. 2005, \apj, 623,
  585

\bibitem[Freytag et al.(1996)]{Freytag:96} Freytag, B., Ludwig,
  H.-G., \& Steffen, M.\ 1996, \aap, 313, 497

\bibitem[Gieren et al.(2015)]{Gieren:15} Gieren, W., Pilecki, B.,
  Pietrzy{\'n}ski, G., et al.\ 2015, \apj, 815, 28

\bibitem[Graczyk et al.(2012)]{Graczyk:12} Graczyk, D., Pietrzy\'nski,
  G., Thompson, I.\ B.\ et al. 2012, \apj, 750, 144

\bibitem[Graczyk et al.(2014)]{Graczyk:14} Graczyk, D., Pietrzy\'nski,
  G., Thompson, I.\ B.\ et al. 2014, \apj, 780, 59

\bibitem[Grevesse \& Sauval(1998)]{Grevesse:98} Grevesse, N., \&
  Sauval, A.\ J. 1998, Space Sci.\ Rev., 85, 161 (GS98)

\bibitem[He{\l}miniak et al.(2015)]{Helminiak:15} He{\l}miniak,
  K.\ G., Graczyk, D., Konacki, M., et al.\ 2015, \mnras, 448, 1945

\bibitem[Herwig et al.(1997)]{Herwig:97} Herwig, F., Bloecker, T.,
  Schoenberner, D., \& El Eid, M.\ 1997, \aap, 324, L81

\bibitem[Iglesias \& Rogers(1996)]{Iglesias:96} Iglesias, C.\ A., \&
  Rogers, F.\ J. 1996, \apj, 464, 943

\bibitem[Sandberg Lacy \& Fekel(2011)]{Lacy:11} Sandberg Lacy, C.\ H.,
  \& Fekel, F.\ C. 2011, \aj, 142, 185

\bibitem[Magic et al.(2010)]{Magic:10} Magic, Z. Serenelli, A., Weiss, 
A. et al.\ 2010, \apj, 718, 1378  

\bibitem[Magic et al.(2015)]{Magic:2015} Magic, Z., Weiss, A., \&
  Asplund, M.\ 2015, \aap, 573, A89

\bibitem[Meng \& Zhang(2014)]{Meng:14} Meng, Y., \& Zhang,
  Q.\ S. 2014, \apj, 787, 127

\bibitem[Michaud et al.(2004)]{Michaud:04} Michaud, G., Richard, O.,
  Richer, J., \& VandenBerg, D.~A.\ 2004, \apj, 606, 452

\bibitem[Moravveji et al.(2015)]{Mora:15} Moravveji, E., Aerts, 
C., P\'apics, P. I., Triana, S. A., Vandoren, B. 2015, \aap, 580, A27

\bibitem[Moravveji et al.(2016)]{Mora:16} Moravveji, E., Townsend,
  R.~H.~D., Aerts, C., \& Mathis, S.\ 2016, \apj, 823, 130

\bibitem[Nieuwenhuijzen \& de Jagger(1990)]{Nieuwenhuijzen:90}
  Nieuwenhuijzen, H., \& de Jagger, C. 1990, \aap, 231, 134

\bibitem[Noels et al.(2010)]{Noels:10} Noels, A., Montalban, J.,
  Miglio, A., Godart, M., \& Ventura, P.\ 2010, \apss, 328, 227

\bibitem[Pavlovski et al.(2014)]{Pavlovski:14} Pavlovski, K., 
 Southworth, J., Kolbas, V., \& Smalley, B. 2014, \mnras, 438, 590

\bibitem[Paxton et al.(2011)]{Paxton:11} Paxton, B., Bildsten, L.,
  Dotter, A.\ et al. 2011, \apjs, 192, 3

\bibitem[Paxton et al.(2013)]{Paxton:13} Paxton, B., Cantiello, M.,
  Arras, P., et al.\ 2013, \apjs, 208, 4

\bibitem[Paxton et al.(2015)]{Paxton:15} Paxton, B., Marchant, P.,
  Schwab, J., et al.\ 2015, \apjs, 220, 15

\bibitem[Pietrzy\'nski et al.(2013)]{Pietrzynski:13} Pietrzy\'nski,
  G., Graczyk, D., Gieren, W.\ et al. 2013, \nat, 495, 76

\bibitem[Pietrzy\'nski et al.(2010)]{Pietrzynski:10} Pietrzy\'nski,
  Thompson, I.\ B., Gieren, W.\ et al. 2010, \nat, 468, 542

\bibitem[Pilecki et al.(2013)]{Pilecki:13} Pilecki, B., Graczyk, D.,
  Pietrzy{\'n}ski, G., et al.\ 2013, \mnras, 436, 953

\bibitem[Pilecki et al.(2015)]{Pilecki:15} Pilecki, B., Graczyk, D.,
  Gieren, W., et al.\ 2015, \apj, 806, 29

\bibitem[Pols et al.(1995)]{Pols:95} Pols, O.\ R., Tout, C.\ A.,
  Eggleton P.\ P., Han, Z. 1995, \mnras, 274, 964

\bibitem[Reimers(1977)]{Reimers:77} Reimers, D. 1977, \aap, 61, 217

\bibitem[Ribas et al.(2000)]{Ribas:00} Ribas, I., Jordi, C., \&
  Gim\'enez, A. 2000, \mnras, 318, 55

\bibitem[Spada et al.(2017)]{Spada:17} Spada, F., Demarque, P., Kim,
  Y.-C., Boyajian, T.\ S., \& Brewer, J.\ M. 2017, \aj, in press
  (arXiv:1703.03975)

\bibitem[Stancliffe et al.(2015)]{Stancliffe:15} Stancliffe, R.\ J.,
  Fossati, L., Passy, J.-C., \& Schneider, F.\ R.\ N. 2015, \aap, 575,
  117

\bibitem[Tkachenko et al.(2014)]{Tkachenko:14} Tkachenko, A.,
  Degroote, P., Aerts, C.\ et al.\ 2014, \mnras, 438, 3093

\bibitem[Torres et al.(2010)]{Torres:10} Torres, G., Andersen, J., \&
  Gim\'enez, A. 2010, \aapr, 18, 67

\bibitem[Torres et al.(2015)]{Torres:15} Torres, G., Claret, A.,
  Pavlovski, K., \& Dotter, A. 2015, \apj, 807, 26

\bibitem[Torres et al.(2014)]{Torres:14} Torres, G., Vaz, L.\ P.\ R.,
  Lacy, C.\ H.\ S., \& Claret, A. 2014, \aj, 147, 36

\bibitem[Valle et al.(2017)]{Valle:17} Valle, G., Dell'Omodarme, M.,
  Prada Moroni, P.~G., \& Degl'Innocenti, S.\ 2017, \aap, 600, A41

\bibitem[Viani \& Basu(2017)]{Viani:17} Viani, L., \& Basu, S. 2017,
  in Seismology of the Sun and the Distant Stars 2016 (Joint TASC2 \&
  KASC9 Workshop and SPACEINN \& HELAS8 Conference),
  eds.\ M.\ J.\ P.\ F.\ G.\ Monteiro, M.\ S.\ Cunha, and
  J.\ M.\ T.\ Ferreira, EPJ Web of Conferences, in press

\end{thebibliography}
\end{document}